\documentclass[conference]{IEEEtran}
\usepackage[pdftex]{graphicx}

\begin{document}

\title{Smart contracts for the Internet of Things: opportunities and challenges}
\author{\IEEEauthorblockN{Nikos Fotiou and
George C. Polyzos 
}
\IEEEauthorblockA{Mobile Multimedia Laboratory, Department of Informatics\\
School of Information Sciences and Technology\\
Athens University of Economics and Business\\
Evelpidon 47A, 113 62 Athens, Greece\\
Email:\{fotiou, polyzos\}@aueb.gr}
}


%


\maketitle

\begin{abstract}
With the Internet of Things (IoT), Things are expected to live in different ``domains'' and ``contexts'' during their lifetime. Information generated by and associated with Things should be manageable by multiple, diverse stakeholders accordingly. Moreover, the scope of the information related to Things can range from private and confidential to public and auditable. Identification, security, and interoperability in this vivid environment are expected to be challenging. In this paper we discuss how smart contracts and blockchain technologies create the potential for a viable solution. To this end, we present smart contract-based solutions that improve security and information management, we identify new opportunities and challenges, and we
provide security recommendations and guidelines.
\end{abstract}

\section{Introduction}
The Internet of Things will create opportunities for new, exciting applications that will interweave the physical with the cyber world. Things will be involved directly or indirectly with the creation of significant amounts of information. A dynamic group of stakeholders should have various levels of access rights on this information. In addition, the scope of the information related to Things will vary depending on the application domain's requirements and the Thing's context. In order to illustrate the diversity of the information associated with or generated by a Thing, as well as the plethora of actions and stakeholders involved in the management of this information, we discuss a use case of a Thing used in a smart building management system. 

The Thing of our use case is capable of performing measurements (e.g., temperature, humidity), it is equipped with a firmware or an operating system for Things (e.g., RIoT~\cite{Bac2013}), and it has connectivity capabilities (e.g., using ZigBee). For each Thing, its manufacturer may provide an ``Application Programming API’’ that can be used by end-users in order to implement new applications and services. This API may leverage a framework, such as Web of Things~\cite{wot} and it may be accessible using connectivity protocol such as CoAP~\cite{rfc7252}). Moreover, the Thing manufacturer may provide a security routine and the necessary (secret) information, that enable Thing update. 
Suppose, the Thing is purchased by a building management company, it is installed in an office, and it is integrated into the building's management system. The building management company now becomes the owner of the Thing, hence it is allowed to extend its API, as well as to associate information with it (such as a CoAP URI). The measurements performed by the Thing can be accessed by various stakeholders, including office residents, the building management company, the city's environment council, and so forth. Moreover, the law specifies that the building should meet some energy efficiency constraints; for this reason, all measurements should be accessible by various auditing agencies. All these stakeholders should be properly authenticated and authorized before performing any interaction with the Thing. In addition, the integrity and the authenticity of the measurements performed by the Thing (as well the other Things), should be protected.

It is clear, even from this simple use case, that IoT applications will involve many stakeholders with different roles, information and functionalities with many access levels, multiple identities and security primitives. Managing all these assets in an efficient, secure, and interoperable manner is a challenging problem. We assert in this paper that the blockchain technology and smart contracts can play an important role towards this direction. 

A blockchain is an append-only ledger of transactions distributed throughout a network of trust-less nodes. Transactions are validated by a number of network nodes (often referred to as the ``miners'') and are added in the ledger upon consensus. Providing that the consensus protocol is \emph{Byzantine Fault Tolerant}, no single entity has control over the ledger. A blockchain may be public~\cite{Wood2014} or restricted to ``permissioned'' users~\cite{hyper}.  Blockchains are considered a ``democratic'' way for maintaining transactions~\cite{ibm2014} and are envisioned to provide novel security mechanisms, to contribute to the sustainability of IoT applications, and to enable new trust models~\cite{Pol2017}.

A smart contract is a distributed app that lives in the blockchain~\cite{Wood2014}. This app is, in essence, a programming language class with fields and methods. Users can interact with the public fields and methods of this class by sending transactions to its ``address'' in the blockchain. Whenever a user interacts with a smart contract, all operations are executed by all nodes in the blockchain network in a deterministic and reliable way; one of these nodes is selected to store the contract’s execution outcome (if any) in the blockchain. Smart contracts can verify blockchain users' identities and digital signatures, they can perform general purpose computations, and they can invoke other contracts. The code of a smart contract is immutable and it cannot be modified even by its owner. Moreover, all transactions sent to a contract are recorded in the blockchain, hence it is possible to obtain all historical values of a contract variable. 
\IEEEpubidadjcol
In the rest of this paper we discuss how smart contracts and the blockchain technology can be leveraged to provide security primitives mechanisms, to facilitate information management, as well as to enable interaction with Things in an interoperable way.   

\section{Information and security management using smart contracts}
\subsection{Smart contract-based security management}
In the following we discuss how common tasks related to security management can be performed using smart contracts.

\textbf{From digital certificates to smart contracts.} All entities interacting with a blockchain (including users and Things) must own at least one public-private key pair. The type of these keys, as well as the signcryption algorithms that can be used with them, are blockchain implementation specific. In any case, these keys can be used as roots of trust. However, building applications using blockchain public keys directly is not practical; means for associating keys with other identifiers and/or attributes (e.g., a human readable name, an application specific identifier, or a QR code) is required. Traditionally, this functionality is implemented using digital certificates. We postulate that smart contracts can play the role of a digital certificate. With this approach, a smart contract can include a list of various forms of identifiers/attributes associated with the public key of the contract owner. Moreover, by making sure that only the contract owner can modify this list,\footnote{This list is a variable, hence it can be modified} and given that the related smart contract lives in the blockchain (hence contract storage and retrieval are inherently protected), additional security guarantees are provided.  Finally, by implementing a resolution mechanism similar to the one described in the following, the need for certificate authorities is negated.

\textbf{Identity space management and resolution.} In contrast to similar approaches (e.g., namecoin~\cite{Kal2015}) we argue that identity assignment should not be implemented following a first-come-first-served approach, instead authorized, well-known entities should vouch for the validity of an identifier. In order to illustrate how this approach can be implemented using smart contracts we discuss the case of DNS. We assume that for the DNS root zone there exist a number of smart contracts and that the addresses of these contracts are well known. Each of these contracts contains mappings from top-level domain names to smart contract addresses. The latter contracts also contain mappings from domain names to smart contract addresses. Finally, the leaf smart contracts contain DNS records. Mappings can be modified only by the appropriate registrants or the domain owners. An interesting property of this system is that all modifications to these mappings are logged and these logs are publicly available. Hence, a registrant cannot modify a mapping to point to another contract without being detected. Moreover, this approach can be used for name resolution: by following the pointers to smart contracts, a user can eventually learn the desired identifier. In a nutshell, we propose that name registrants should maintain their role in managing who can register which identity, but all other processes should be implemented in a blockchain using smart contracts.  Although, as we discuss in Section IV, this solution is not recommended for storing mappings from DNS names to IP addresses, it is ideal for storing mappings from DNS names (or other forms of identifiers) to security related (and other) information.

\textbf{Facilitating legacy and novel security mechanisms}. Smart contracts that can be retrieved by an identity resolution mechanism (similar to the one described  previously) may store additional security primitives, such as TLS public keys (see for example Fig. 1). By storing keys in the blockchain, users do not have to rely on pre-trusted certificate authorities (as for example in the Web Public Key Infrastructure~\cite{Dur2013}). Moreover, due to the blockchain transparency and distributed nature, this approach is more robust against malicious authorities, compared for example to the DANE protocol~\cite{Hof2012}, which relies on trusted DNS servers and registrants. Additionally, these smart contracts may as well contain other auxiliary information that can be used by contemporary security mechanisms (for example~\cite{Fot2016} uses the blockchain to store  the ``System Parameters’’ of an Identity-Based encryption scheme~\cite{Gre2007})

\subsection{Smart contract-based information management} 
Smart contratcs and the blockchain technology can faciliate and improve information management in many ways. 

\textbf{Access Control.} The modification of the variables of a smart contract and the execution of its methods can be easily restricted to certain authorized (blockchain) users. Furthermore, smart contracts can be used for authenticating a user. For example, a smart contract may include a method that raises an `event' whenever it is invoked: every time an entity receives this event it can trivially obtain the (blockchain) public key of the user that invoked the corresponding method. This authentication mechanisms can be a component of an access control solution.  For more advances access control solutions,  blockchain-based secure messaging technologies--such as Catenis~\cite{cate}--can be used for exchanging \emph{authorization tokens} and smart contracts can be used for verifying tokens' validity~\cite{Zys2015}. Finally, access control decisions can be logged to the blockchain, facilitating this way auditing and accountability mechanisms.

\textbf{Information and functionality scoping.}  An interesting property of smart contracts is that multiple users can deploy the same contract: each instance of the contract will have its own address and it will be executed independently of the others. This property facilitates information and functionality scoping. Imagine for example a Thing manufacturer that creates a device. The manufacturer can create a stub smart contract that ``describes'' the device (an interface in terms of object-oriented programming), as well as client applications/libraries that can interact with this contract. A third party can extend the contract provided by the manufacturer, by adding the
desired functionality to the corresponding methods, and deploy it: existing applications/libraries will still be able to interact with this new contract, providing that they are configured with its address.

\textbf{Proof of ownership, information verification, non-repudiation.} A common problem of today's information systems is that it is often hard to prove the ownership of an information item. Similarly, it is hard for a user to prove that he interacted with a particular instance of an information item (for example, he invoked a CoAP URI and he received a specific measurement), but on the other hand it is easy for a malicious user to present a fake ``proof'' of an interaction with a specific information instance (for example, a modified log file). The blockchain technology enables information owners to ``announce'' an item (hence have a proof of ownership), as well as communicating endpoints to agree upon and share transcripts of interactions. Even more interestingly, this blockchain-assisted logging can be implemented as an extension to existing security protocols. For example the solution presented in~\cite{Rit2017} extends TLS to create proofs about the content of a TLS session which are then uploaded in a blockchain. Additionally, and since blockchains are append-only ledgers, we can achieve non-repudiation, i.e., it is not possible for users to claim that they did not approve a transaction.

\begin{figure}
\begin{center}
\includegraphics[width=0.95\linewidth]{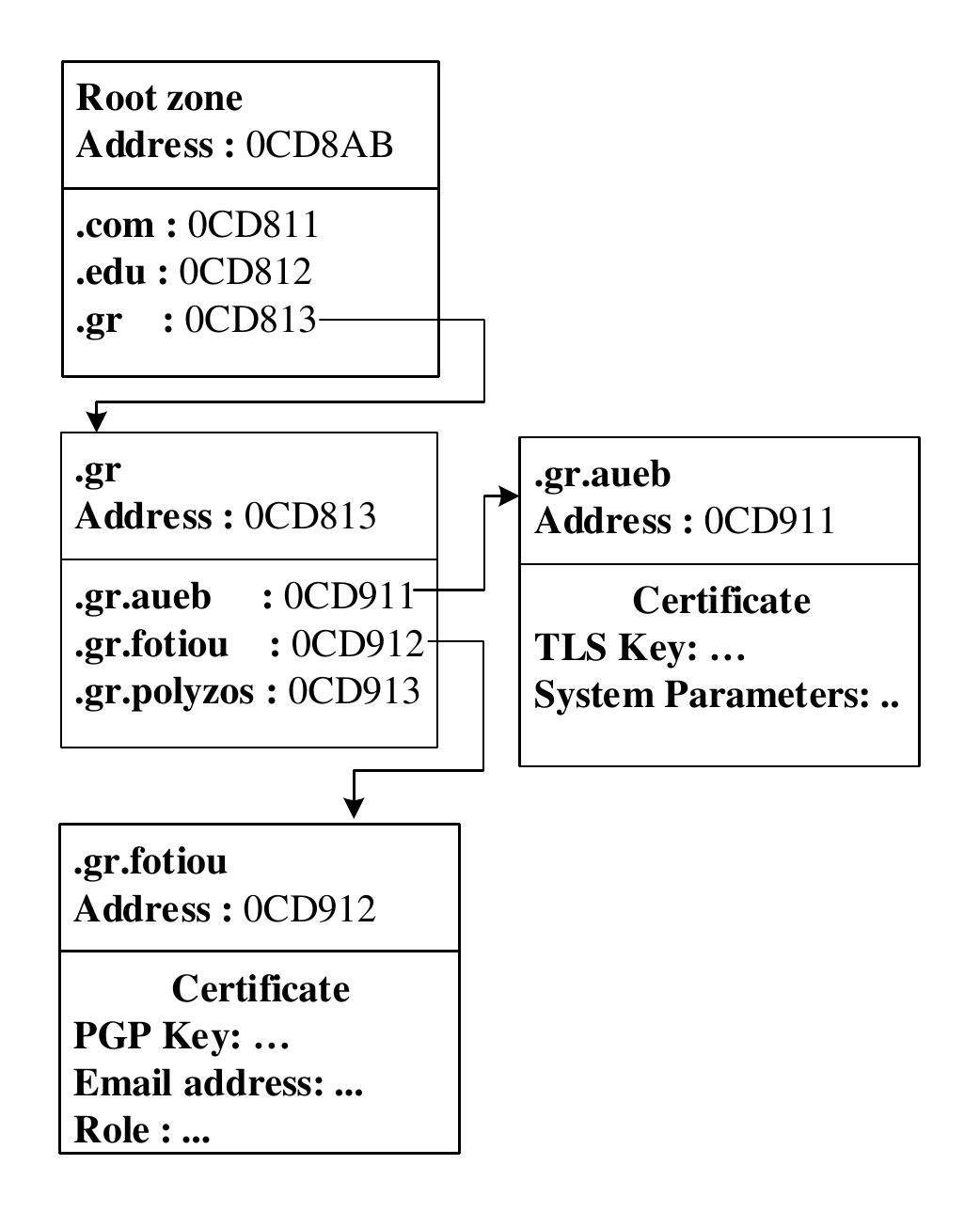}
\caption {Public key resolution for the domain aueb.gr using smart contracts. The address of the smart contract that represents the root zone is considered well-known.}
\label{fig:namespace}
\end{center}
\end{figure}  

\section{Interacting with Things using smart contracts}
We envision smart contracts enabling applications to interact with Things in the IoT, in a way similar to how hardware drivers enable applications to interact with hardware devices; i.e., smart contracts can describe the capabilities of a Thing, the services it offers, as well as how it can be accessed. By extending existing smart contracts, developers should be able to integrate Things into their systems and processes, as well as to offer innovative, sustainable services. Moreover, by leveraging the trust anchors provided by smart contracts and the underlay blockchain, it should be easier to build open, reliable, and secure distributed applications, as well as novel security, accountability, and charging mechanisms. Similarly, with smart contracts users can easily perform transactions with their digital currencies, or even their custom tokens. 

An interesting property of smart contracts is that they are deterministic and they are always executed ``correctly,'' therefore, a contract owner should not worry whether or not the application logic (e.g., an access control condition) included in the contract will be ``respected.'' Moreover, the code of a contract is generally available and a contract owner may even add additional checks in a contract in order to protect its ``customers'' even from himself! For example, he can implement a two-phase commit for payments: a costumer instead of paying a service provider directly, he commits some digital money/tokens to a contract; these funds are held in escrow by the contract until the service provider provides the expected service; in case he fails to do so, the funds are returned to the customer. Finally, smart contracts cannot be removed from a blockchain. This property is useful for building censorship resistant communication mechanisms. For example, and as we discuss in the following, a smart contract can provide a pointer to the location of an information item (or even the item itself): if this location becomes (somehow) blacklisted, the pointer can be easily updated by the contract owner. Similarly, a smart contract may contain meta-data that can be used for preventing frauds and fake content items.

In the following, we discuss some smart contract-based interaction models for the IoT. It should be noted that when we state Things interact with a blockchain or a smart contract, it is always implied that this is through a (trusted) gateway. Moreover, the term users in the following may refer to real world entities, or applications, or Things, or even smart contracts.

\subsection{Push-Pull}
With this interaction model, Things make data (content, information) available (i.e., they push it) and users can then pull it. For the push operation the following approaches can be considered:
\begin{itemize}
\item Things push data in the blockchain. With this approach, data is stored in the blockchain (possibly through a smart contract) and smart contracts can be used for retrieving it. Storing data in the blockchain may result in scalability issues and it may raise some security concerns (we discuss security issues in the following section). However, there are some cases where this is necessary, e.g., in applications where full transparency is required, or in applications where data should be audited by third parties. Moreover, smart contracts may provide some short of aggregation and data analytics services. For example, a smart contract may aggregate temperature measurements and provide methods that provide the last measurement, or a daily average/min/max, etc.

\item Things push data onto storage nodes. With this approach, Things store data in dedicated storage nodes. Then they store a ``pointer'' to that node and auxiliary meta-data in the blockchain through a smart contract. Depending on the form of this pointer, a user may pull data using various methods. For example, if this pointer is an HTTP URL the user should perform an HTTP request; if this pointer is an address in the blockchain, the user should pull content by issuing a transaction, etc. The auxiliary meta-data may be information that can be used for the verification of the content integrity (e.g., a hash of the content data), information that can be used for verifying content provenance and authenticity (e.g., a digital signature), access control related information, content description, etc. 
\end{itemize}

\subsection{Publish-Subscribe}
With this interaction model, users express interest in a data item (subscribe) and Things send data items (publish) to the interested users. The subscription process can be implemented using a smart contract, which should maintain a list of ``pointers'' to interested users. Each smart contract can be responsible for a specific topic. Moreover, topic hierarchies can be considered allowing wildcard subscriptions. A resolution mechanisms, similar to the one described previously--or even an of-chain directory, should be used for mapping a topic to a smart contract address. 

Things should monitor topics for new subscriptions (alternatively, a notification service can be implemented). Every time a new item becomes available, it should be published to all its subscribers. The publication process implementation depends on the form of the used ``pointers to subscribers'', i.e., it can be implemented as a transaction in the blockchain, or as an HTTP operation, etc.

\subsection{Event-based}
A form of interaction in the IoT is ``actuation.'' Performing an actuation through a smart contract is not trivial, since contracts do not interact with the physical world. On the other hand, some smart contract implementations provide ``events.'' Using events, actuation can be implemented as follows: actuation operations could be implemented as smart contract methods, users could invoke these methods, which in return can create an event; Things could monitor the blockchain for events and if an event takes place, they would perform the appropriate actuation.

\section{Security and privacy considerations} 
Compared to traditional Web applications, smart contracts have some particularities that should be considered when designing a distributed application. 

\textbf{Smart contracts are open}. A core principle of blockchain-based systems is transparency: everybody can access all records stored in the blockchain, and since smart contracts are part of the blockchain, anybody can view the ``source code'' of a smart contract, or even execute it offline.  Therefore, smart contracts should not implement methods and algorithms which should be kept secret.

\textbf{Information related to smart contracts is always available}. In addition to the code of a smart contract, all users of a blockchain are able to view the values that contract variables hold, historical data, as well as, all transactions related to that contract. This property has many implications: (i) smart contracts cannot store private data (e.g., private keys, protected records), (ii) smart contracts cannot perform operations that require secret information (e.g., create a digital signature), (iii) smart contracts cannot generate secret information (e.g., generate a secret key). Additionally, this property creates privacy concerns: smart contracts should not be used for storing sensitive data (e.g., medical records). This also applies to encrypted data: since data cannot be removed from the blockchain, the discovery of a security flaw may enable attackers to decrypt this data at some point in the future. Even the fact that somebody interacts with a smart contract may constitute a privacy threat, for example interacting with a smart contract that sells medicines may reveal some information about your health. Finally, this property enables security attacks that were not possible using alternative, legacy systems. For example, consider the identity management approach described in the previous section and suppose it was used for mapping DNS records to IP addresses: by reading the smart contract, a malicious user would be able to enumerate all IP addresses of a target. 

\textbf{Smart contracts are immutable}. Once deployed, smart contracts cannot be modified (in some cases they can only be ``killed''). Therefore, if there is a flaw in the contract's application logic, this flaw will exist forever, since there is no way to provide updates. For this reason, and as we discuss in the next section, smart contracts should not be overwhelmed with code, instead ``indirections'' should be used.  

When it comes to the IoT, there are some additional considerations that should be taken into account. Things do not have the necessary computational power to interact with a blockchain, in many cases they do not even have the power to act as a ``light node,'' that is a node that only sends and receives transactions to the blockchain, but it does not store the blockchain and it does not perform any form of verification. For this reason, a gateway-based approach should be used. The gateway should store Things' secret information and act on behalf of the Things. 

Smart contracts cannot interact with external services, let alone the physical world. For this reason, creating two-phases commit that involve, e.g., an actuation, is not a trivial task.

\section{Discussion and design guidelines}
Smart contracts facilitate business development since they provide a convenient and transparent mean for offering secure services. The fact that smart contract execution is deterministic and nobody, not even the contact owner, can affect the execution output creates potential for new, innovative applications. On other hand, the immutability and the transparency of smart contracts create serious risks, since a flaw in the contract's application logic may result in funds being stolen. For example, a flaw in a smart contract of the DAO distributed app, resulted in \$156M loss~\cite{Har2016}, leading eventually to the creation of a new ``fork'' of the ethereum blockchain. For this reason, we propose a skeleton-based design approach for smart contract-based distributed apps. With this approach, there is a core contract, the skeleton contract, which defines all methods. However, instead of including the actual implementation for each method, the skeleton contract provides a pointer to other contracts, which implement the defined method. Hence, a smart contract method execution is a two-step transaction: firstly, the user performs a transaction with the skeleton contract and retrieves the address of the contract implementing the desired method, and secondly the user performs a transaction with the latter contract invoking the desired method. If a flow is detected in a method implementation, then a new contract for this method should be created and the corresponding pointer in the skeleton contract should be updated. The skeleton contract must implement a check that will verify that the user who updates a pointer is the contract owner.

As already stated, the contract source code, as well as the value of each contract field is publicly available. For this reason contract developers should be really careful when deciding what information will be stored in a contract. In many cases it is preferable to store data outside the blockchain and store in the blockchain the hash of the data. This way, a user can verify the integrity of the received data. Moreover, and for the same reason, developers are advised to use tools that automatically analyze contract code and detect possible flaws and security risks (for example~\cite{Kal2018}).

Finally, contract developers should consider a ``kill switch'', i.e., a piece of code (often provided by the as a service by various smart contract providers) that will render a contract useless. Invoking a kill switch should result in all funds associated with the contract being transferred to the contract owner and in all users being prevented from interacting with this contract anymore. It should be noted that even if a contract is killed, its code and data will remain in the blockchain.   

\section{Conclusions} 
Smart contracts are a new, exciting tool that creates new potentials for the IoT. Smart contracts have some unique features that create opportunities for novel, secure applications. Such features are: they are transparent, they are executed in a deterministic way by third parties and nobody can affect their execution output, they provide means for user authentication and token transfer, all interactions with a smart contract are logged in the blockchain. On the other hand, smart contracts are not a panacea since they come with risks and weaknesses: once deployed they cannot be modified, they do not preserve user privacy, and they cannot store or create secret information. 

In this paper, we postulated that smart contracts can be used as an abstraction that will connect applications with Things, interweaving eventually the Internet of Things. To this end, we discussed security and information management solutions using smart contracts, we presented design choices for user-Thing interaction, we discussed security considerations, and we presented design guidelines. 

We believe that we are still missing an efficient and secure way for gluing Things with smart contracts, since Things do not have the necessary computational power for interacting with the blockchain. Moreover, and by taking into consideration that research efforts on Things access protocols (e.g., CoAP) and on blockchain access and interoperability protocols (e.g., W3C's Interledger) are advancing in parallel (and orthogonally from each other), we envision a gateway-based protocol handler, which will translate from IoT-specific protocols into blockchain transactions, and vice versa, in an efficient and secure way.

\bibliographystyle{IEEEtran}
\bibliography{IEEEabrv,diss}
\end{document}